# Ballistic Imaging of High-Pressure Fuel Sprays using Incoherent, Ultra-short Pulsed Illumination with an Ultrafast OKE-based Time Gating

Harsh Purwar[1,*], Saïd Idlahcen[1], Claude Rozé[1], Jean-Bernard Blaisot[1]

1: CORIA UMR-6614, Normandie Université, CNRS, Université et INSA de Rouen, St. Etienne du Rouvray, France
* correspondent author: harsh.purwar@coria.fr

**Abstract**  We present an optical Kerr effect based time-gate with the collinear incidence of the pump and probe beams at the Kerr medium, liquid carbon disulfide ($CS_2$), for ballistic imaging of the high-pressure fuel sprays. The probe pulse used to illuminate the object under study is extracted from the supercontinuum generated by tightly focusing intense femtosecond laser pulses inside water, thereby destroying their coherence. The optical imaging spatial resolution and gate timings are investigated and compared with a similar setup without supercontinuum generation, where the probe is still coherent. And finally, a few ballistic images of the fuel sprays using coherent and incoherent illumination with the proposed time-gate are presented and compared qualitatively.

## 1. Introduction

Over the past few years, due to technological advancements in the laser light sources and detection devices, there has been a lot of efforts in developing optical diagnostic tools for applications in various fields of science and technology, since most of these tools are non-intrusive in nature. However, many real-world applications and phenomena of interest are intrinsically linked to turbid environments where light scattering and attenuation strongly limit the interpretation of optical signals. In a wide range of applications, from imaging in biological tissues [1,2], to measurements of high-pressure multiphase flows involving cavitation or turbulent breakup [3], key information is scrambled by the distortion imparted to the light signal as it transits the measurement volume. Informative optical diagnostics in such media require detailed understanding of the light source, propagation and scattering in the measurement volume, and a detection arrangement tailored to collect the meaningful parts of the transmitted light signal.

Assuming a vanishingly short input pulse and a ~1 cm turbid measurement volume, the typical transmitted pulse durations can be expected to be of the order of 50-100 ps, with most of the informative signal present in the first few picoseconds of the transmitted signal [3]. Consequently, most time gating applications require arrangements that can limit light collection to a few picoseconds or shorter time window.

To this end, ultrafast time gating can provide an effective means of segregating high integrity portions of the collected signal from light disturbed by scattering interactions. On an average, photons that participate in more interaction events traverse a more circuitous path through the medium and are statistically more likely to be distorted or redirected from their original trajectories. This difference in the optical path length results in a temporal spreading of the light. Time gating, or time filtering allows selection of the signal that retains the information on the object characteristics.

However, a common problem experienced by researchers trying to develop an ultrafast time-gated imaging technique for ballistic imaging or time-resolved imaging, in general, is due to the illumination source itself, where an ultra-short laser pulse is used as imaging or probe pulse to illuminate the object under study. The unavoidable artifacts arising due to the speckles, produced by the illumination source itself and the multiple-diffraction patterns, provided path length differences are comparable to the wavelength of the incident laser beam, as is in high-pressure fuel sprays, could be partly removed if the coherence of the laser beam is destroyed. This is important because even with the modern sophisticated image processing tools, analysis and interpretation of such images, in terms of morphology or velocity estimation for example, are then very complicated. As is shown in the later sections of this proceeding, the images obtained with the incoherent illumination are much more cleaner and sharper compared to the images obtained with direct laser illumination.

Another issue arises due to the non-collinear overlap of the pump (or switching beam) and the probe beam.

In the classical OKE-based time-gates the angle between the pump and the probe beams not only influences optical gate's temporal characteristics but also is the reason for a major drawback of the non-collinear optical time-gates [4]. Due to the angle between the pump and the probe beams, the overlap between them does not occur at the same time. As a result the obtained time-resolved images do not correspond to the same time from the left to the right side of the image. Here, we present an approach with collinearly overlapping pump and probe beams for ultrafast time-gated imaging, using an incoherent ultra-short source derived from supercontinuum (SC), generated using a femtosecond laser, for illumination. Introducing an ultra-short incoherent illumination source avoids these artifacts keeping the pulse width short enough as demanded by the gating applications and the collinear overlap takes care of most of the issues with the OKE-based time gating as are discussed in the later sections.

## 2. Experimental Setup

A schematic of the experimental setup for ballistic imaging of high-pressure fuel sprays, using an incoherent, ultra-short pulsed light source for illumination followed by an ultrafast optical Kerr effect based time-gating is shown in Fig. 1.

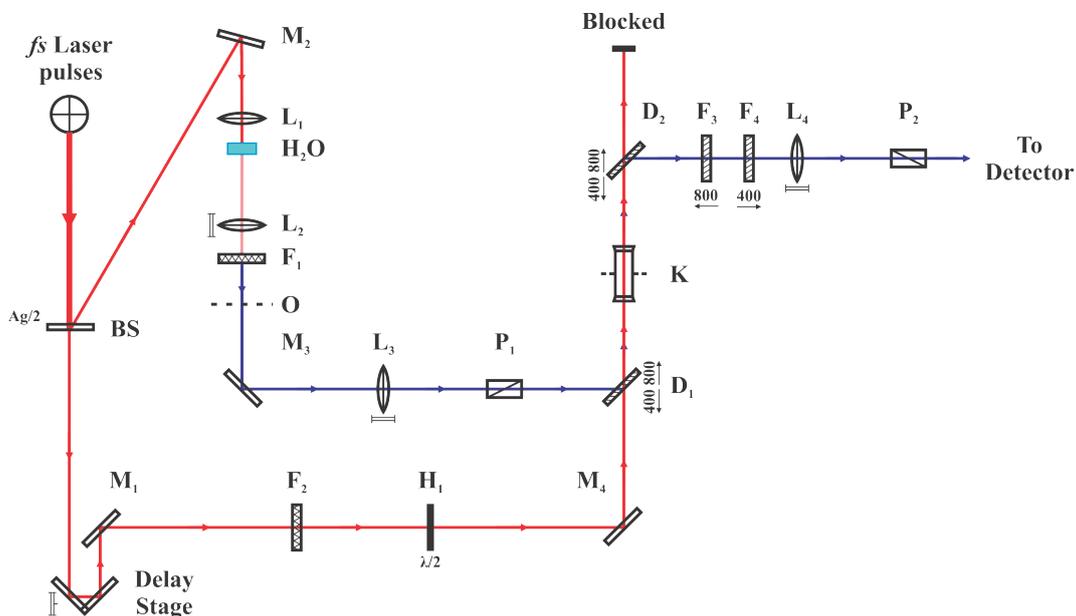

**Fig. 1** Schematic of the experimental setup for ballistic imaging with incoherent, ultra-short supercontinuum-derived pulsed illumination and collinear incidence of the pump and probe beams inside the Kerr medium.

The femtosecond laser pulses were generated using a regenerative amplifier (Libra, Coherent) seeded with a Titanium-Sapphire mode-locked laser (Vitesse, Coherent). The system is capable of generating laser pulses with pulse width less than 120 fs and energy of about 3.5 mJ per pulse at a repetition rate of 1 kHz, pulse spectrum centered at wavelength $\lambda$ = 800 nm. The incoming fs laser beam was separated into probe and pump beams using a 50:50 beam splitter (BS). The pump pulse ($\lambda$ = 800 nm), used to activate Kerr effect in the Kerr medium (K) for a duration depending on the non-linear properties of the medium and on the pump pulse characteristics, passes through a computer controlled delay stage with a least count of 0.1 μm (corresponding to a temporal delay of 0.67 fs in air). The delay between the pump and the probe pulses is adjusted so that both pulses arrive at the Kerr medium at the same time and a good alignment takes care of their spatial overlap. A neutral density filter ($F_2$) is used to adjust the power of the pump beam and a half wave plate ($H_1$) to rotate its polarization axis by an angle, with respect to the probe polarization, which maximizes the induced birefringence in the Kerr medium, thus maximizing the efficiency of the time-gate.

The probe pulse is extracted from the supercontinuum generated by tightly focusing the high power femtosecond laser pulses inside a 5 cm wide cuvette filled with distilled water [5-7] using a short focal length biconvex lens $L_1$. An appropriate band pass filter ($F_1$, $\lambda$ = 330-480 nm) extracts a narrow range of wavelengths for illumination of the fuel spray and at the same time also suppresses the incoming fundamental wavelength ($\lambda$ = 800 nm). An appropriate notch filter could also be used to block the incoming fundamental wavelength if necessary. It is well known that the pulse width for individual wavelengths after supercontinuum generation does not change significantly [8] even if the overall pulse width may increase as light propagates due to the dispersion. The band pass filter also reduces this artifact. The probe pulse then illuminates the object (O) under study.

Since the pump and the probe beams here have different wavelengths, it is possible to combine them before the Kerr medium (K) and to separate them afterwards using the dichroic mirrors ($D_1$ and $D_2$). Using this principle both pulses are incident at the Kerr medium in a collinear fashion. After interaction with the object, the probe pulse passes through the Kerr medium, which is sandwiched between two crossed polarizers $P_1$ and $P_2$. A pair of band pass filters ($F_3$ and $F_4$) may be used to filter out the pump beam reaching the detector due to its scattering by the Kerr medium. An additional neutral density filter may be used to adjust the intensity of light reaching the detector according to its limitations. For imaging purposes a CCD Camera (Hamamatsu C9100-02) was used as a detector whereas for basic intensity measurements a power meter (Ophir Nova-II) was used. A fiber-optic spectrometer (Ocean Optics Maya2000 Pro) was used for basic spectral measurements.

Note that because the supercontinuum generation is due to the combined effect of many coupled nonlinear processes such as filamentation, self-phase modulation, self-focusing/defocusing etc. inside water, the coherence of the laser beam is destroyed and thus in the time-filtered images obtained with this technique, unwanted diffraction patterns and laser speckles are reduced. Also since the pump and probe beams can now be separated easily using notch and band-pass filters the noise due to the high power pump beam is highly reduced in this collinear configuration of the OKE-based time-gate.

## 3. Results and Discussion

The spectrum of the supercontinuum generated using 800 nm intense laser beam ranges from visible to near infrared region (400 – 1100 nm) as can be seen in Fig. 2. A notch filter was used to cut off the fundamental wavelength ($\lambda = 800$ nm) from the spectrum. A small band of wavelength ($\lambda = 330 - 480$ nm) was selected from this continuum to be used as the probe beam for illumination of the object under study and is highlighted in red in Fig. 2.

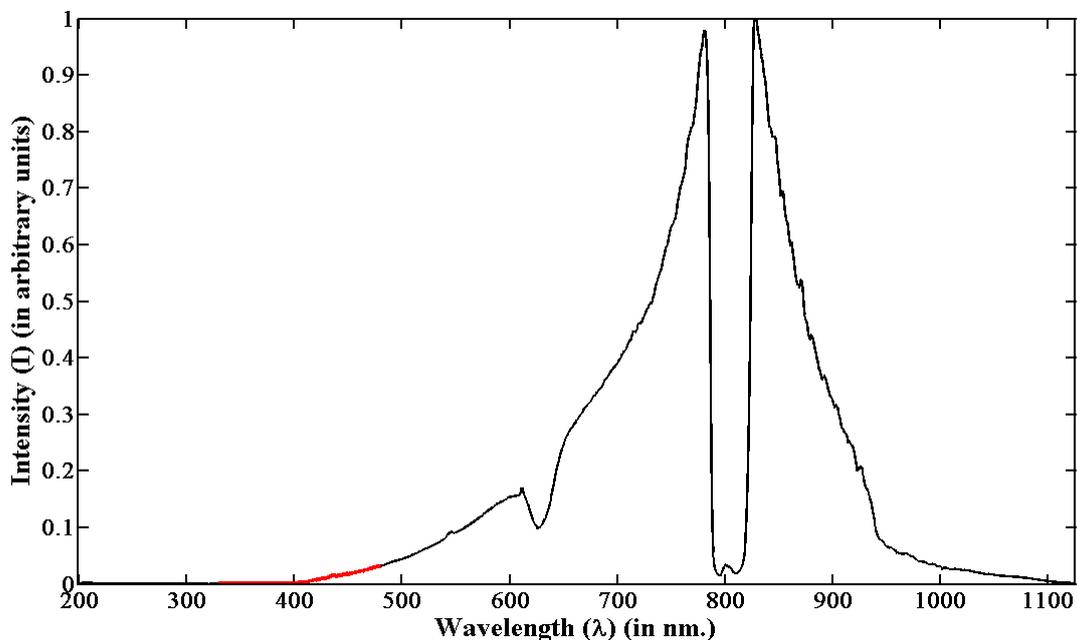

**Fig. 2** Supercontinuum spectrum and the selected wavelength range used for illumination of the object under study (curve highlighted in red, $\lambda = 330\text{-}480$ nm).

The first simple experimentation without optical time gating showed, as was expected, that the quality of the images is highly improved by using incoherent illumination: Fig. 3 shows a magnified view of a section from the diesel spray images obtained directly by illuminating the spray (a) with the coherent laser pulse and (b) with the incoherent light pulse extracted from the SC in the similar external conditions of pressure and temperature. The injection pressure for both these cases was constant at 400 bars. Clearly, the image with the incoherent illumination is free from the unwanted artifacts of the multiple diffractions and laser speckles. Also note that the duration of the continuum pulse is short enough to freeze the motion.

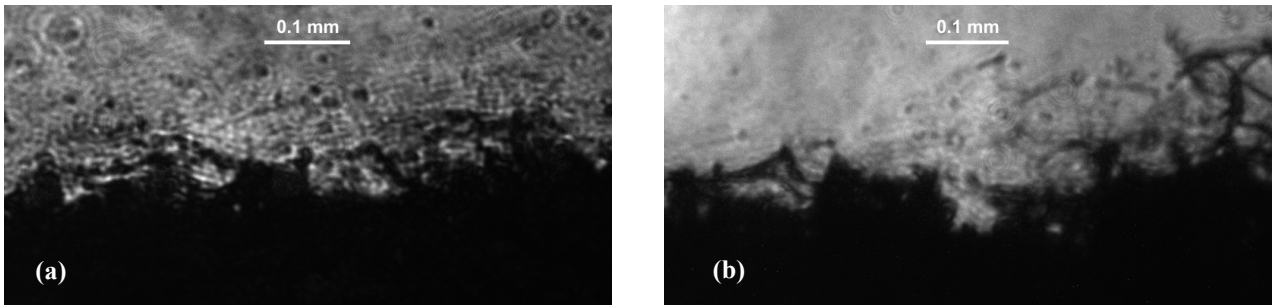

**Fig. 3** A zoomed section of the spray images obtained using (a) direct laser and (b) supercontinuum derived illumination for qualitative comparison.

**Characterization of the optical gate:**
The temporal resolution of the optical gate with a supercontinuum-derived probe beam ranging from λ = 330 – 480 nm is shown in red in Fig. 4. This graph was extracted from the averaged spectra measured using the aforementioned spectrometer by varying the delay between the pump and the probe pulses using the delay stage as shown in Fig. 1. The extended or delayed response of the Kerr medium for the SC-derived probe (right side of the peak) is due to the dispersion. The same is not observed when the temporal profile is measured for a fixed wavelength of light.

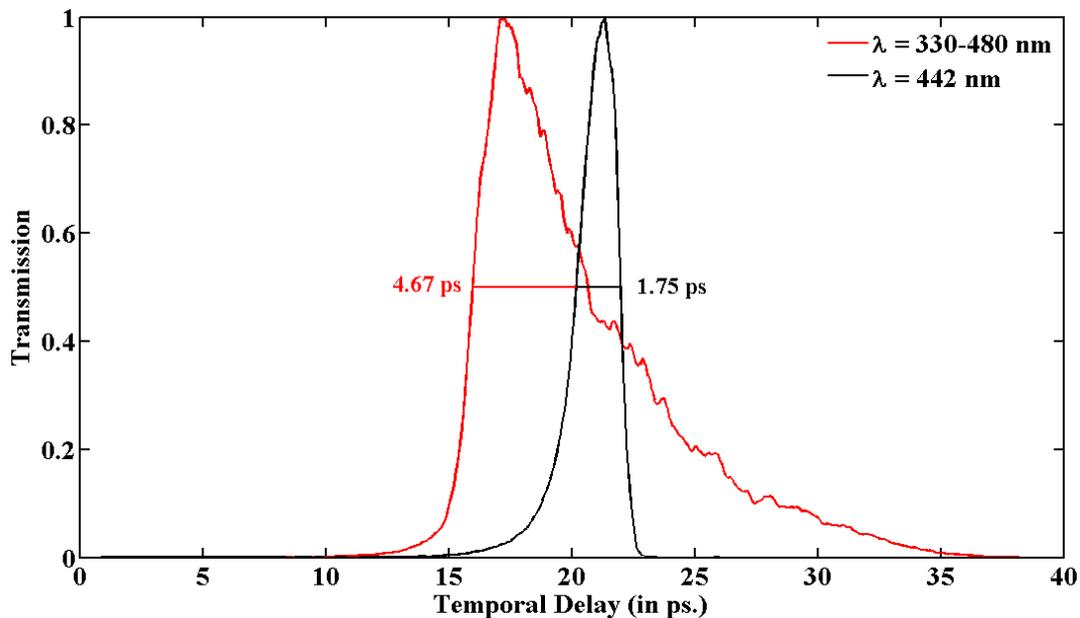

**Fig. 4** The temporal profile of the OKE-based time gate (1.0 mm $CS_2$ as Kerr medium) for SC-derived probe pulse (red) and for a particular wavelength (λ = 442 nm, in black) extracted from the same SC.

The spatial resolution is estimated using the normalized modulation transfer function (MTF), which, for the presented optical setup (Fig. 1) was measured using the slanted-edge method (ISO 12233 standard for MTF measurement of electronic still-picture cameras [9]). The MTF and edge-spread function (ESF), $F(x)$ are related as follows,

$$MTF(\nu) = \int \frac{dF(x)}{dx} \exp(-2\pi i\nu x)\, dx$$

where $\nu$ is the spatial frequency in lines/mm, $x$ is the abscissa along an axis orthogonal to the slanted edge. In other words, MTF is the Fourier transform of the derivative of the ESF, which is also the line-spread function. It should be noted that while calculation of MTF from the over-sampled ESF data, the differentiation along the slanted edge is very sensitive to the high-frequency noise and in order to avoid amplifying this noise during the computation of the derivative, the averaged over-sampled ESF is fitted using the Logistic (Fermi) function, $f(x)$ [10],

$$F(x) = d + f(x) = d + \frac{a}{1 + \exp[(x-b)/c]}$$

where a, b, c and d are the desired fit parameters.

Figure 5 and 6 show the fitted ESF and normalized MTF calculated for the SC-derived probe with its collinear overlap with the pump inside the Kerr medium (1.0 mm liquid $CS_2$) placed at the image plane of the lens $L_3$. For comparison Fig. 6 also shows the normalized MTF without generating supercontinuum but using a single wavelength for the probe beam (λ = 400 nm) obtained by second harmonic generation in a beta-barium borate (BBO) crystal [11]. The achievable imaging spatial resolution from the MTF curve (inverse of Nyquist frequency) with 1.0 mm liquid $CS_2$ as the Kerr medium placed at the image plane of $L_3$ for the supercontinuum-derived probe was found to be 147.6 lines/mm while for the frequency-doubled probe in the same configuration it was found to be 93.2 lines/mm.

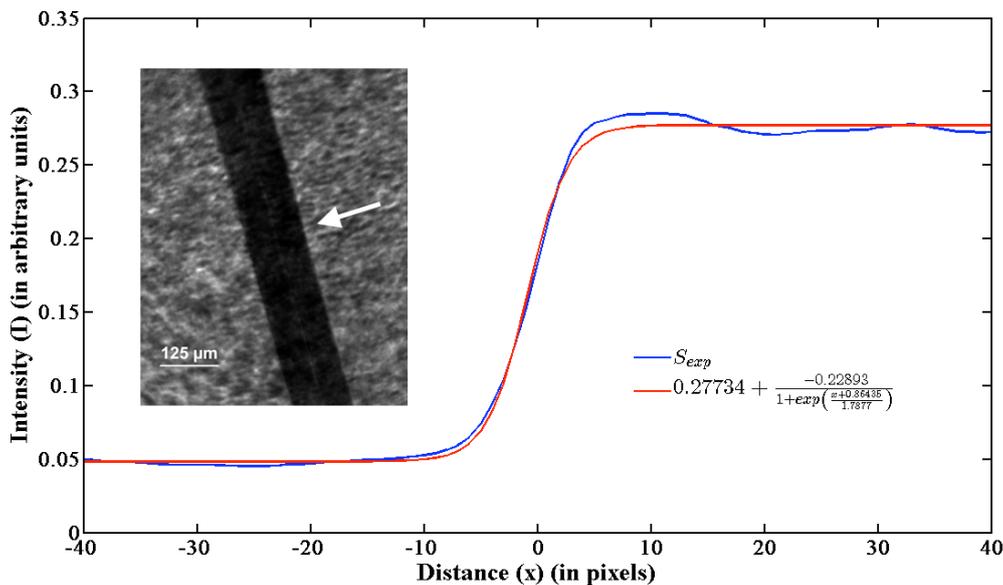

**Fig. 5** Image of a slanted glass fiber (ϕ = 125 μm) obtained using SC-derived illumination with collinear optical time-gate configuration and the corresponding averaged edge-spread function (ESF) for the indicated edge in the image.

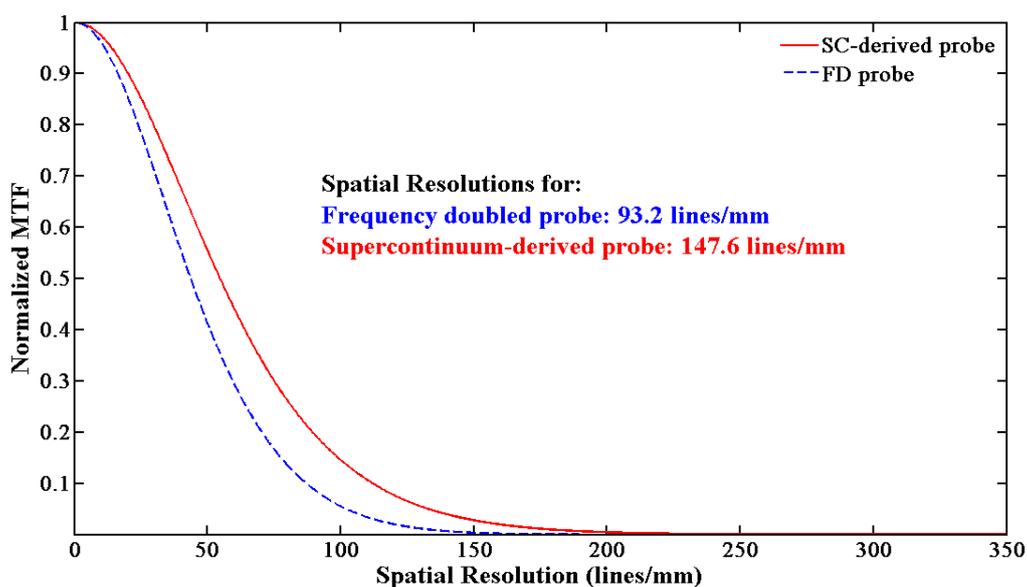

**Fig. 6** The normalized modulation transfer function (MTF) as a function of spatial frequency (ν) (in lines/mm) calculated using the fitted ESFs for the collinear configuration of the optical gate with SC-derived (red, solid line curve) and frequency doubled (blue, dotted line curve) probe beams for illumination of the object under study.

And finally after the successful development and characterization of the optical gate using probe pulse derived from SC, it has initially been tested for the imaging of high-pressure fuel sprays. Figure 7 shows a comparison of the spray images 5 cm far from the nozzle (single-orifice injector, nozzle diameter = 185 µm) obtained with the collinear configuration of the optical gate with coherent (frequency doubled) and incoherent (SC-derived) illumination sources. The injection pressure was kept constant at around 400 bars for both the cases. The effect of speckles is clearly visible in the time-filtered images obtained by using coherent illumination whereas the same is largely reduced when the spray is illuminated by the supercontinuum derived, incoherent light.

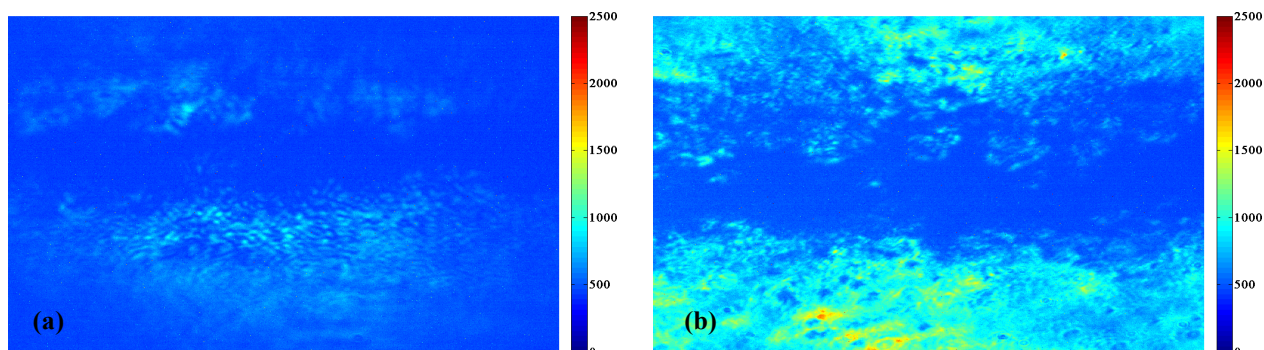

**Fig. 7** Ballistic images of fuel spray with (a) frequency-doubled, coherent and (b) SC-derived, incoherent illumination sources with the collinear OKE-based time-gate.

## 4. Conclusions

An OKE-based time-gate is presented with the collinear incidence of the pump and the probe beams at the Kerr medium, 1.0 mm liquid $CS_2$, thereby solving several issues linked with the non-collinear overlap of the two in the classical configuration of the optical gate. In order to remove some artifacts arising due to the coherence of the laser beam in the spray images, for example laser speckles and diffraction patterns, the coherence of the laser beam was destroyed by tightly focusing it inside a water cuvette generating a wide band of white light continuum. A narrow band of wavelengths were selected from this supercontinuum for illumination of the sprays instead. Even though the gate duration for this supercontinuum derived probe beam has increased to about 4.5 ps, the quality of the time-resolved images has improved significantly due to the incoherent illumination. Preliminary results for the spray images obtained using this incoherent, ultra-short, pulsed illumination with the collinear OKE-based time filtering have been presented.